\documentclass[runningheads]{llncs}
\usepackage[T1]{fontenc}
\usepackage[misc]{ifsym}
\usepackage{graphicx}
\usepackage{amssymb}
\usepackage{makecell}
\usepackage{algorithm}
\usepackage{algorithmic}
\usepackage{amsmath,amsfonts}
\usepackage{multirow}
\usepackage{diagbox}
\usepackage[symbol]{footmisc}
\usepackage{subfigure}
\usepackage[colorlinks,
            linkcolor=blue,       
            anchorcolor=blue,  
            citecolor=blue, 
            urlcolor=blue,
            ]{hyperref}

\begin{document}
\title{Enhancing Privacy of Spatiotemporal Federated Learning against Gradient Inversion Attacks}
\titlerunning{Spatiotemporal Gradient Inversion Attack}
%
\author{Lele Zheng\inst{1,2} \and Yang Cao\inst{1}\textsuperscript{(\Letter)} \and Renhe Jiang\inst{3} \and Kenjiro Taura\inst{3} \and Yulong Shen\inst{2} \and  Sheng Li\inst{4} \and Masatoshi Yoshikawa\inst{5}}
\authorrunning{L. Zheng et al.}
%
\institute{Tokyo Institute of Technology, Tokyo, Japan \email{llzhengstu@gmail.com, cao@c.titech.ac.jp} \and
Xidian University, Xi’an, China \email{ylshen@mail.xidian.edu.cn}\\ \and The University of Tokyo, Tokyo, Japan \email{jiangrh@csis.u-tokyo.ac.jp, tau@eidos.ic.i.u-tokyo.ac.jp} \\ \and National Institute of Information and Communications Technology (NICT), Kyoto, Japan \email{sheng.li@nict.go.jp}\\ \and Osaka Seikei University, Osaka, Japan \email{yoshikawa-mas@osaka-seikei.ac.jp}}
\maketitle              
\begin{abstract}
Spatiotemporal federated learning has recently raised intensive studies due to its ability to train valuable models with only shared gradients in various location-based services.
On the other hand, recent studies have shown that shared gradients may be subject to gradient inversion attacks (GIA) on images or texts. However, so far there has not been any systematic study of the gradient inversion attacks in spatiotemporal federated learning. In this paper, we explore the gradient attack problem in spatiotemporal federated learning from attack and defense perspectives. To understand privacy risks in spatiotemporal federated learning, we first propose Spatiotemporal Gradient Inversion Attack (ST-GIA), a gradient attack algorithm tailored to spatiotemporal data that successfully reconstructs the original location from gradients. Furthermore, we design an adaptive defense strategy to mitigate gradient inversion attacks in spatiotemporal federated learning. By dynamically adjusting the perturbation levels, we can offer tailored protection for varying rounds of training data, thereby achieving a better trade-off between privacy and utility than current state-of-the-art methods. Through intensive experimental analysis on three real-world datasets, we reveal that the proposed defense strategy can well preserve the utility of spatiotemporal federated learning with effective security protection.
\end{abstract}

\section{Introduction}\label{sec:intro}
Spatiotemporal data analysis tasks, such as human mobility predictions (HMP), play a pivotal role in various fields due to their potential to predict and analyze movement patterns of individuals or groups~\cite{zhang2023federated}. Accurate predictions of human mobility can enhance urban planning~\cite{yuan2012discovering}, recommend appropriate points of interest~\cite{liu2017point}, and facilitate transportation systems in smart cities~\cite{shi2019survey}. Due to privacy concerns, federated learning (FL) has been extensively employed in human mobility prediction, enabling clients to train shared models collaboratively without directly disclosing their private data~\cite{feng2020pmf,li2020predicting}. 
\begin{figure}[ht]
    \centering
    \includegraphics[width=0.95\textwidth]{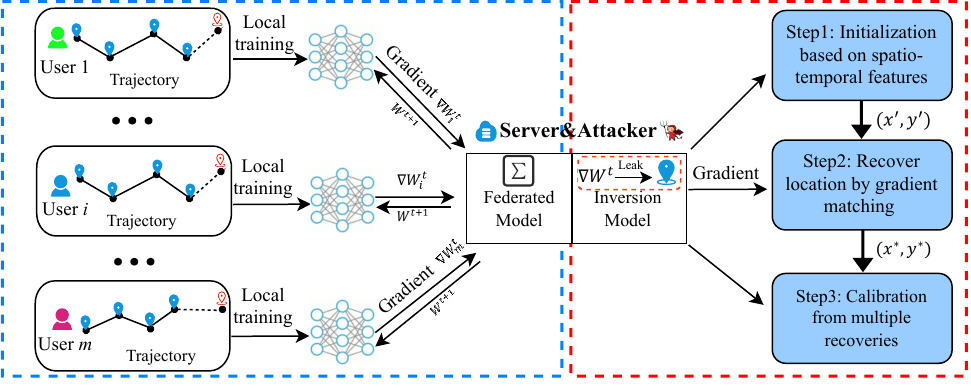}
    \caption{Overview of ST-GIA. The left part (blue box) performs the federated protocol, and the right part (red box) illustrates the main steps of ST-GIA.}
    \label{fig: system}
\end{figure}

Although federated learning inherently enhances privacy by enabling clients to keep private data on their local devices, recent studies have highlighted the potential vulnerability of shared gradients to gradient inversion attacks~\cite{fan2024guardian,geiping2020inverting,geng2023improved,zhu2019deep}. As shown in Fig.~\ref{fig: system}, while ostensibly adhering to protocol, the honest-but-curious server might covertly steal privacy by leveraging gradient inversion attacks on gradients shared by clients, thereby reconstructing raw data. Most gradient inversion attacks primarily focus on minimizing the distance between the dummy gradients and the ground-truth gradients. To generate dummy gradients, random data and corresponding labels are fed to the global model. Taking the distance between the gradients as error and the dummy inputs as parameters, the recovery process can be formulated as an iterative optimization problem. Upon the convergence of this optimization process, the private data is expected to be comprehensively reconstructed. Several differential privacy-based methods have been proposed to preserve privacy in spatiotemporal federated learning, such as DPSGD~\cite{abadi2016deep}, GeoI~\cite{andres2013geo}, and GeoGI~\cite{takagi2020geo}.

However, although the existing gradient inversion attacks have undeniable contributions, they still have some limitations that cannot be ignored. (1) First, the existing attacks are developed to reconstruct training \textit{images} or \textit{texts} used to train classifiers and have yet to be validated in spatiotemporal federated learning. (2) Second, current gradient inversion attacks tend to analyze each round in isolation, overlooking the integration of vital information from the overall training process. Within the context of spatiotemporal federated learning, the outcome of an attack in a given round is likely to influence the attack result in the subsequent round significantly. (3) Third, in the context of spatiotemporal federated learning, achieving effective outcomes through the direct application of existing differential privacy-based defense methods proves challenging. The reason is that these methods are primarily designed for general-purpose defenses and are not tailored to address gradient leakage attacks specifically.

In this paper, we propose a novel gradient attack algorithm, named ST-GIA, designed explicitly to raise awareness of spatiotemporal data, which can effectively reconstruct the original location from shared gradients. In ST-GIA, the attackers first exploit the characteristics of spatiotemporal data to initialize dummy data. Then, the attackers recover the original location through gradient matching, effectively leveraging a priori knowledge of the road network to improve the accuracy of their attacks significantly. Finally, they employ multiple recovery results to refine and calibrate the final reconstruction outcomes.  In addition, we evaluate the effectiveness of existing differential privacy-based defense methods against ST-GIA and propose a new adaptive privacy-preserving strategy tailored to mitigate gradient inversion attacks in spatiotemporal federated learning. In particular, we design an importance-aware budget allocation method to ensure the sensitivity to different training rounds. The main contributions are summarized as follows:
\begin{enumerate}
    \item As the first attempt in the field of spatiotemporal federated learning, we propose ST-GIA, a gradient attack algorithm for spatiotemporal data that can effectively reconstruct the original location from shared gradients.
    \item We design an adaptive privacy-preserving strategy tailored to mitigate gradient inversion attacks in spatiotemporal federated learning. This strategy leverages the attack risk to measure the privacy sensitivity of various training model rounds and to allocate privacy budgets adaptively.
    \item Comprehensive experiments conducted on three real-world datasets demonstrate the effectiveness of the proposed attack and defense strategies.
\end{enumerate}

\section{Preliminaries}
\subsection{Spatiotemporal Federated Learning}
One typical task requiring spatiotemporal federated learning is human mobility prediction, which utilizes historical trajectories  $s_u = \{x_u^{0}, x_u^{1}, \dots, x_u^{n} \}$ to predict the location $x_u^{n+1}$ of the target user $u$ in the next time step, 
where spatiotemporal point $x_u={(t, lat, lon)}$ can be described as a 3-tuple of the time stamp, latitude, and longitude. In such a scenario, spatiotemporal federated learning aggregates model parameters from different clients into a global model, where the clients learn the temporal and spatial correlations of the data locally. 

\textbf{Local Training at a Client.} The client first downloads the global state $w_t$ from the server and then performs local training using the local trajectory data, i.e., $w_{t+1}=w_t-\eta\nabla w_t$, where $w_t$ is the local model parameter update at round $t$ and $\nabla w_t$ is the gradient of the model parameters. Several models can be used for local training to reveal spatiotemporal correlations in trajectory data, such as LSTM~\cite{hochreiter1997long}, DeepMove~\cite{feng2018deepmove}, and so on.

\textbf{Update Aggregation at FL Server.} Upon receiving the local updates from clients, the server aggregates these updates to get a global state. In this paper, we evaluate the widely adopted federated averaging algorithm~\cite{mcmahan2017communication}, where the server iteratively updates the global model by computing a weighted average of the incoming weight parameters.

\subsection{Gradient Inversion Attack}
A gradient inversion attack is carried out by the server, or an entity that has compromised it, with the intention of acquiring a client's private data $(x^*, y^*)$ through the analysis of gradient updates 
$\nabla_{\theta^k}\mathcal{L}(x^*, y^*)$ uploaded to the server. These attacks typically presume the presence of honest-but-curious servers, which adhere to the federated training protocol without altering it. The common method for obtaining private data is to solve an optimization problem:
\begin{equation}
    \arg\min\limits_{(x', y')} \delta(\nabla_{\theta^k}\mathcal{L}(x^*, y^*), \nabla_{\theta^k}\mathcal{L}(x', y')),
\end{equation}
where $\delta$ represents a specific distance measure and $(x', y')$ denotes dummy data. Typical choices for $\delta$ are $L_2, L_1$, and cosine distances. 

\subsection{Local Differential Privacy}
Local differential privacy (LDP)~\cite{bassily2015local} has emerged as the gold standard for protecting individual privacy in scenarios where user data is collected by an untrusted data collector. Essentially, LDP enables users to determine the extent to which their data is distinguishable to the data collector through a privacy parameter, $\epsilon$, chosen by the user.
\begin{definition}[Local Differential Privacy]
A randomized algorithm $\mathcal{M}$ satisfies $\epsilon$-local differential privacy if for any two inputs $x, x'\in \mathcal{D}$ and for any output $y\in \mathcal{Y}$, the following equation holds:
\begin{equation}
{\rm Pr}[\mathcal{M}(x)= y]\leq {\rm exp}(\epsilon)\cdot{\rm Pr}[\mathcal{M}(x')= y],
\end{equation}
\end{definition}
A smaller $\epsilon$ guarantees stronger privacy protection because the adversary has lower confidence when trying to distinguish any pair of inputs $x, x'$.

\begin{definition}[Sensitivity]
For any pair of neighboring inputs $d, d' \in \mathcal{D}$, the sensitivity $\Delta f$ of query function $f(\cdot)$ is defined as follows:
    \begin{equation}
        \Delta f = \max\limits_{d, d'} \left\|f(d) - f(d')\right\|_1,
    \end{equation}
    where the sensitivity $\Delta f$ denotes the maximum change range of function $f(\cdot)$.
\end{definition}

\begin{definition}[Exponential Mechanism]
    Given a score function $q: (\mathcal{D}, y) \rightarrow \mathcal{Y}$, a random algorithm $\mathcal{M}$ satisfies $\epsilon$-differential privacy, if
\begin{equation}
    \mathcal{M}(\mathcal{D},q)=\left\{y:\vert {\rm Pr}[y\in \mathcal{Y}] \varpropto {\rm exp}(\frac{\epsilon(\mathcal{D}, y)}{2\Delta q})\right\},
\end{equation}
where $\Delta q$ is the sensitivity of the score function $q: (\mathcal{D}, y) \rightarrow \mathcal{Y}$.
\end{definition}

\begin{definition}[Constrained domain]
    We denote $\mathcal{C}^t = \{x_i| {\rm Pr}(x^t = x_i)>0, x_i \in \mathcal{X}\}$ as constrained domain, which indicates a set of possible locations at $t$, where $x^t$ is the user's true location at $t$ and $x_i\in \mathcal{X}$.
\end{definition}

\section{Spatiotemporal Gradient Inversion Attack}\label{Spatiotemporal Gradient Inversion Attack}
Previous studies have delved into reconstructing input data from gradients; see Table~\ref{comparison_state_of_art}. However, current works focus on reconstructing training images or texts and have yet to be validated in spatiotemporal federated learning. Our initial evaluation of these methods in the context of spatiotemporal federated learning revealed that, while some methods can partially recover real data, the success rate is generally low (see Section~\ref{PERFORMANCE EVALUATION} for more details). We attribute this limited effectiveness to the lack of consideration of spatiotemporal features in existing methods. We propose ST-GIA, a novel gradient attack algorithm tailored to spatiotemporal data that effectively reconstructs the original location from gradients. As shown in Fig.~\ref{fig: system}, it consists of three main steps: initialization based on spatiotemporal features, recovering location by gradient matching, and calibration from multiple recoveries.

\begin{table}[t]
\caption{Comparing different gradient inversion attacks in spatiotemporal federated learning. We show the success rates of various attacks on the NYCB dataset, where the local training model employs an LSTM architecture (ASR: attack success rate, which is defined in Section~\ref{PERFORMANCE EVALUATION}). }
\centering
\small
\begin{tabular}{|c|c|c|c|c|c|}
    \hline
      Method       &  Optimization terms  & Initialization  & Model &  \makecell{Additional}& ASR \\
      \hline
      DLG \cite{zhu2019deep}  & $l_2$ distance &  Gaussian   & LeNet&  -   & 0.434 \\
      \hline
      iDLG \cite{zhao2020idlg}  & \makecell{$l_2$ distance} &Uniform &LeNet& -  & 0.290\\
      \hline
      InvGrad \cite{geiping2020inverting} & \makecell{Cosine similarity} &Gaussian & ResNet&  TV norm  & 0.013 \\
      \hline
      CPL \cite{wei2020framework} & \makecell{$l_2$ distance} &Geometric &LeNet &   \makecell{label based \\ regularizer} & 0.321\\
      \hline
      SAPAG \cite{wang2020sapag}& \makecell{Gaussian kernel \\based function} & Constant &ResNet &  -& 0.125  \\
      \hline
       \textbf{ST-GIA} (ours)& $l_2$ distance & ST-based &ST models & \makecell{Mapping \\ Calibration} & 0.652  \\
      \hline
    \end{tabular}
\label{comparison_state_of_art}
\end{table}

\subsection{Initialization based on Spatiotemporal Features} 
To reconstruct the spatiotemporal data, we first initialize dummy data, denoted as $(x', y')$, where $x'$ represents the dummy input and $y'$ is the dummy label. Subsequently, we derive the corresponding dummy gradient as follows:
\begin{equation}
    \nabla w^{\prime} \leftarrow \partial \mathcal{L}(F(x^{\prime},w_t),y^{\prime})/\partial w_t.
\end{equation}

There are different strategies for initializing the dummy data. Among these, random Gaussian noise is the most commonly utilized technique for data initialization in image and text recovery tasks. In addition, constant values or random noise sampled from Uniform distribution are also presented for data initialization. Jonas et al.~\cite{geiping2020inverting} have shown that gradient attacks frequently fail to achieve convergence due to poor initialization in image reconstruction scenarios. Our experimental findings also confirm that improper initialization adversely impacts the effectiveness of attacks in spatiotemporal federated learning. Consequently, we propose that attackers can effectively use spatiotemporal data characteristics for dummy data initialization. Specifically, an attacker might leverage the reconstructed location from a previous round as a basis to initialize the dummy point in the subsequent round. This approach takes advantage of the continuity inherent in user mobility and makes an educated guess. Therefore, for the attack in the $t$-th round, we can initialize the dummy data as follows:
\begin{equation}
    x^{\prime t} \gets x^{\prime t-1}, 
    y^{\prime t} \sim \mathcal{N}(0,1).
\end{equation}
In the case of attacking the first training data round, we continue to employ random Gaussian noise for the initialization of dummy data.

\subsection{Gradient Matching}
The next step involves optimizing the dummy gradient, $\nabla w^{\prime}$, to closely approximate the ground truth gradient, $\nabla w$. To achieve this, we must define a differentiable distance function, $D(\nabla w^{\prime}, \nabla w)$, enabling us to determine the optimal $x'$ and $y'$, denoted as $(x^*, y^*)$, as follows:
\begin{equation}
    (x^*, y^*)= \arg\min\limits_{(x', y')} D(\nabla w^{\prime}, \nabla w).
\end{equation}

\textbf{Distance Function.} We consider the $l_2$ norm (Euclidean distance) as our distance function to measure the difference between $\nabla w^{\prime}$ and $\nabla w$.  The Euclidean distance fits the characteristics of the spatiotemporal data since it is a natural metric of distance between locations.
\begin{equation}
    D(\nabla w^{\prime}, \nabla w) = \left \| \nabla w^{\prime}-\nabla w \right\|_2.
\end{equation}

\textbf{Mapping.} We focus on the task of predicting human mobility, enabling us to utilize prior knowledge of the road network to enhance attack accuracy. We assume that all location coordinates are on the road network, which is a reasonable assumption since many human mobility prediction tasks are based on the road network. Therefore, within the attack optimization process, if the outcome of an attack iteration falls outside the road network, we map the position $x'$ to its nearest point on the network. The ablation experiment in Section~\ref{PERFORMANCE EVALUATION} shows that this simple operation can greatly increase the attack success rate. This improvement is attributed to helping the attack model converge and ensuring that the attack results for each iteration remain within a plausible range.

\begin{algorithm}[t]
\caption{ST-GIA}\label{Algorithm 1}
\begin{algorithmic}[1] 
\REQUIRE{$F(x; w_t)$: Differentiable machine learning model, $\nabla w_t$: model gradients after target trains at round $t$, learning rate $\eta$ for optimizer, $N$: max attack iterations. $T$: global training rounds.}
\ENSURE{reconstructed training data $(x^{*},y^*)$}
\FOR{$t=1$ to T }
\IF{$t=1$}
\STATE Initialize dummy locations $x_0^{\prime} \sim \mathcal{N}(0,1), y_0^{\prime} \sim \mathcal{N}(0,1)$.
\ELSE
\STATE Initialize dummy locations $x_0^{\prime t} \gets x_{N+1}^{\prime t-1}, y_{0}^{\prime} \sim \mathcal{N}(0,1)$.
\ENDIF
\FOR{$i \leftarrow 1$ to $N$}
  \STATE $\nabla w_i^{\prime} \leftarrow \partial \mathcal{L}(F(x_i^{\prime},w_t),y_i^{\prime})/\partial w_t$
    \STATE $\mathbb{D} \leftarrow \left \| \nabla w_i^{\prime} - \nabla w  \right \|^2$
    \STATE $x_{i+1}^{\prime} \leftarrow x_i^{\prime} - \eta\nabla_{x_i^{\prime}}\mathbb{D}_i, y_{i+1}^{\prime} \leftarrow y_i^{\prime} - \eta\nabla_{y_i^{\prime}}\mathbb{D}_i$
    \STATE \textit{Mapping} $(x_{i+1}', y_{i+1}' )$
    \ENDFOR
\ENDFOR
\STATE \textit{Calibration} from multiple recoveries according to Equation \ref{Calibration}.
\STATE \textbf{return} $(x^{*}, y^*)$ 
\label{DLG}
\end{algorithmic}
\end{algorithm}

\subsection{Calibration from Multiple Recoveries} 
The gradient is uploaded in each interaction round to capture the spatiotemporal relationships among data points. As a result, each data point undergoes multiple training iterations. Consequently, for a given data point, the attacker obtains multiple reconstruction results. When the global training round $t$ is less than the timesteps ($T_s$) of training, each point is reconstructed $t$ times. Conversely, if the number of rounds exceeds $T_s$, reconstruction occurs $T_s$ times for each point. Inspired by this finding, the attacker can aggregate multiple reconstruction results to yield a more precise outcome. Simplified, the average of all reconstructed locations for data point can be considered the final attack result:

\begin{equation}
x^{*} =
\begin{cases}
\frac{1}{|T_s|} \sum_{T_s} x', & \mbox{if } t \ge |T_s| \\
\frac{1}{|t|} \sum_{t} x', & \mbox{if } t \leq |T_s|
\end{cases}    
\label{Calibration}
\end{equation}
It can be considered as a group consistency from multiple recoveries that helps get closer to the global optimal point.

\subsection{The Framework of the Algorithm}
Algorithm~\ref{Algorithm 1} demonstrates our proposed ST-GIA. In each global round, we initialize dummy data $(x_0', y_0')$ based on spatiotemporal features in lines 2–6. We obtain the dummy gradient $\nabla w^{\prime}$ corresponding to the dummy input in line 8. We then use Euclidean distance to measure the distance between the dummy and true gradients (line 9). After each attack iteration, we update $(x_i', y_i')$ in line 10. If $(x_i', y_i')$ are not within the range of the road network, then we map $(x_i', y_i')$ to the location on the road network closest to it. When the preset maximum number of iterations $N$ is reached or the dummy data no longer changes, we obtain the preliminary attack result. After obtaining all reconstruction results of a location, we calibrate all preliminary results to obtain the final reconstructed position $(x^{*}, y^*)$.
\begin{figure}[htbp]
\centering
\begin{minipage}[t]{0.48\textwidth}
\centering
\includegraphics[width=5cm,height=4cm]{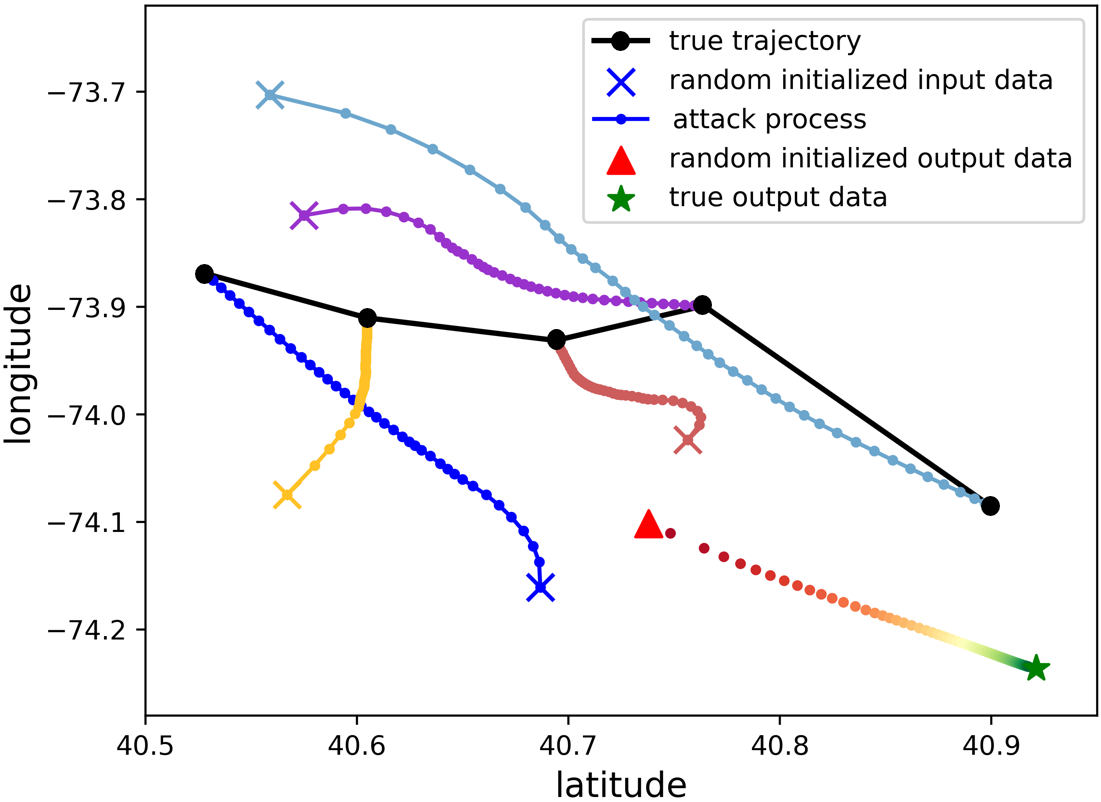}
\caption{A reconstructed trajectory}
\label{A reconstructed trajectory.}
\end{minipage}
\begin{minipage}[t]{0.48\textwidth}
\centering
\includegraphics[width=5cm,height=4cm]{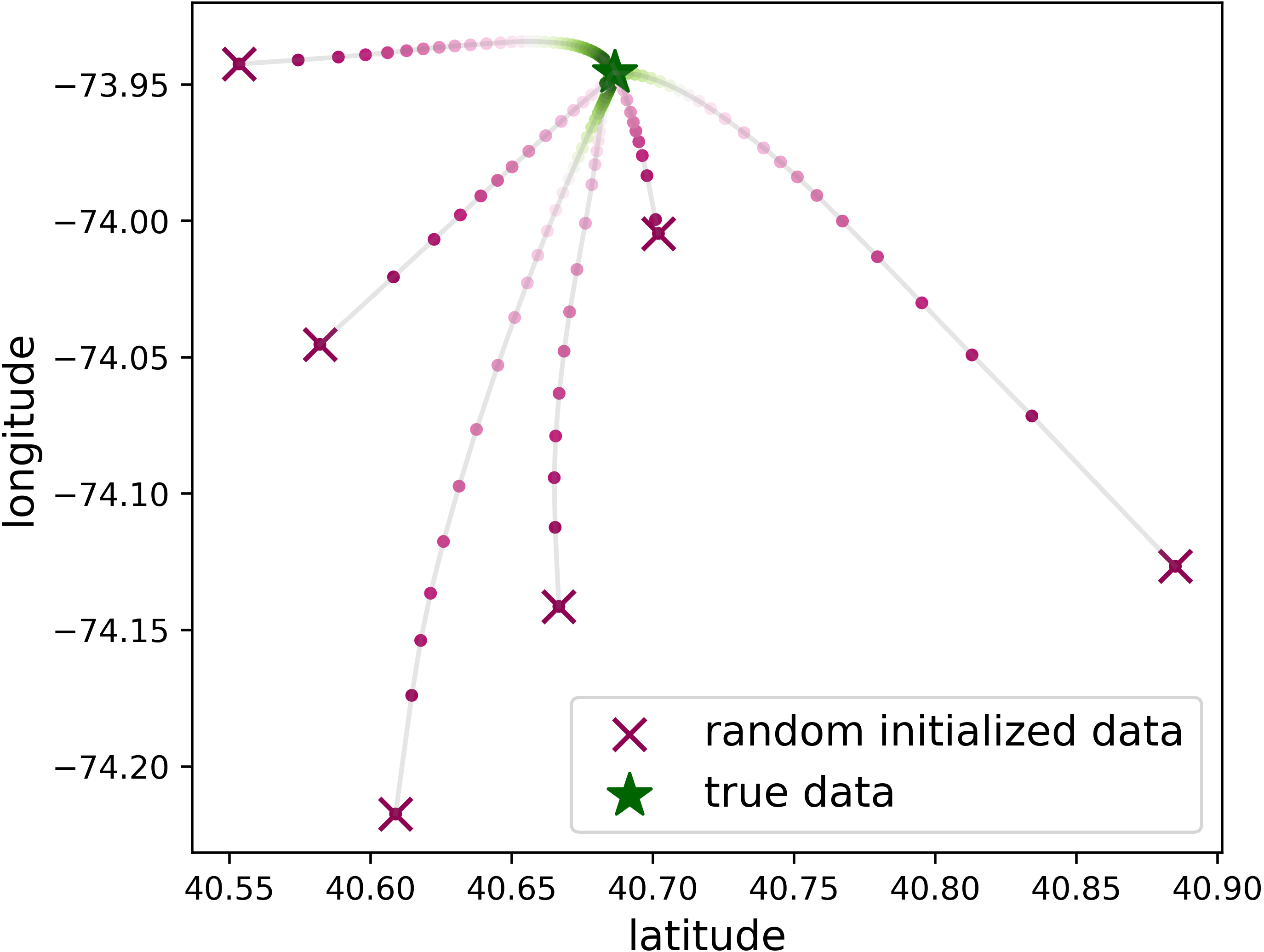}
\caption{Different initialization}
\label{Different initialization.}
\end{minipage}
\end{figure}

Fig.~\ref{A reconstructed trajectory.} illustrates the performance of an attacker to reconstruct a user’s trajectory utilizing ST-GIA, wherein each reconstructed trajectory point closely approximates the corresponding true trajectory point. We present the attack process on the first-epoch global model, thus employing random initialization. Upon convergence of the attack model, the minimum distance between the reconstructed and true locations is reduced to a mere 1 meter. Moreover, we find that the attack results converge in the direction of the true location, regardless of the initialization location. We conduct multiple experiments under the real-world dataset and found that to be true, independent of the initialization dummy location; see Fig.~\ref{Different initialization.}.

\section{Adaptive Privacy-preserving Strategy}
Provable differential privacy may remain the only way to guarantee formal privacy against gradient inversion attacks. We evaluate three differentially private methods on three real-world datasets, as described in Section~\ref{PERFORMANCE EVALUATION}. We find that achieving effective outcomes by directly applying existing differential privacy-based defense methods proves challenging. The reason is that these methods are primarily designed for general-purpose defenses and are not tailored to address gradient inversion attacks specifically. We design an adaptive privacy-preserving strategy to mitigate gradient inversion attacks in spatiotemporal federated learning, which mainly consists of adaptive budget allocation and perturbation based on personalized constraint domains.

\subsubsection{Adaptive Budget Allocation.} We have observed that the reconstruction error increases as the global model converges, which can be attributed to the attacker needing the information leaking from the gradient to reconstruct the user’s location. Consequently, the privacy level encountered in various rounds is not uniform. However, the traditional approach assumes that each round in the training process is equally important; therefore, the privacy budget is allocated equally to each round. Such an allocation often gives rise to issues, notably the overprotection of certain rounds and the underprotection of others.

\begin{algorithm}[t]
\small
\caption{Adaptive privacy-preserving strategy.}\label{Algorithm 2}
\begin{algorithmic}[1] 
\REQUIRE{$\epsilon$: the total privacy budget, $T$: max global training rounds, $\mathcal{C}_k^t$: the constraint domain of user $k$ at $t$, $x_k^t$: input data for user $k$ in round $t$.}
\ENSURE{Model parameter $\hat{w_k^t}$.}
\FOR{$t = 1$ to $T$}
    \STATE $\epsilon'=\epsilon-\sum_{i=1}^{t-1}\epsilon_j$
    \STATE $\epsilon_t = exp(-\gamma[t]) \cdot \epsilon'$
    \FOR{each user $k$ \textit{in parallel}}
        \STATE $\hat{x_k^t}=PGEM(\mathcal{C}_k^t, \epsilon_t, x_k^t)$
        \STATE $\hat{w_k^t} \leftarrow LocalUpdate(k, w_{t-1}, t, \hat{x_k^t})$
        \STATE \textbf{return} $\hat{w_k^t}$ to server 
    \ENDFOR
\ENDFOR
\label{Adaptive privacy-preserving strategy}
\end{algorithmic}
\end{algorithm}

Inspired by this idea, we propose an adaptive privacy budget allocation method. Specifically, our main idea is that during the training process, users should dynamically adjust the perturbation of their local data in response to the varying importance of distinct rounds. The assessment of this importance depends on the attack risk. We use two metrics, Attack Success Rate (ASR) and Attack Iteration (AIT), to measure the attack risk. ASR reflects the proportion of training data accurately reconstructed in a particular round to the total training data in that round. Obviously, a higher attack success rate indicates that the attacker can get more accurate data. We also employ attack iteration (AIT) to measure the cost of the attack, where AIT is the number of attack iterations required for a successful attack. A lower AIT value signifies greater efficiency for the attacker. Thus, the importance of round $t$ can be computed as follows:
\begin{equation}
    \gamma[t]=\alpha\mathcal{F}_1(ASR[t])+\frac{\beta}{\mathcal{F}_2(AIT[t])},
\end{equation}
where the functions $\mathcal{F}_1(\cdot)$ and $\mathcal{F}_2(\cdot)$ denote the effect of ASR and AIT on importance, respectively. The parameters $\alpha, \beta$ are weight factors and $\alpha + \beta = 1$.

We aim to provide adaptive protection for the training data. Rounds with a higher risk require intensified security measures, i.e., more noise should be added. Therefore, the proportional function that decides the portion of the remaining budget allocated to the current round can be defined as:
\begin{equation}
    p=exp(-\gamma[t]).
\end{equation}
The exponential function guarantees that $p$ ranges from 0 to 1. The final budget allocated to the current round is
\begin{equation}
    \epsilon_i = p \cdot \epsilon',
\end{equation}
where $\epsilon'$ is the remaining budget $\epsilon' = \epsilon-\sum_{i=1}^{t-1}\epsilon_t$.

\subsubsection{Personalized Constraint Domain.} 
To ensure privacy, each user can define a personalized constraint domain based on their individual requirements; for example, a student may be active only on campus, then she can define her own constraint domain as all locations on the entire campus. Personalized constraint domain can be denoted $\mathcal{C}_k^t = \{x_i^k| {\rm Pr}(x_k^t = x_i)>0, x_i \in \mathcal{X}\}$, which indicates a set of possible locations of user $u_k$ at $t$.

We propose an obfuscation mechanism, PGEM, considering each user's personalized constraint domain so that it can output more useful locations. PGEM uses the idea of an exponential mechanism~\cite{4389483} to perturb the true location of user $u_k$. Given the input $x\in \mathcal{X}$, the privacy budget $\epsilon_t$, the constraint domain $\mathcal{C}_k^t$, PGEM outputs $c\in \mathcal{C}_k^t$ with the following probability:

\begin{equation}
   { \rm Pr[PGEM }(x) = c] =\frac{e^{-\frac{\epsilon_t}{2}d(x, c)}}{\sum_{c\in \mathcal{C}_k^t}e^{-\frac{\epsilon_t}{2}d(x, c)}}{,}
     \label{PGEM}
\end{equation}
where $d(\cdot)$ is the distance metric between two locations. Simply, we can use Dijkstra's algorithm, which means that  $d(x, \cdot)= Dijkstra(\mathcal{C}_k^t, x)$.
\subsubsection{Adaptive Privacy-preserving Strategy.} We propose an adaptive privacy-preserving mechanism to protect location privacy in spatiotemporal federated learning. Our algorithm is shown in Algorithm~\ref{Algorithm 2}. It consists of three phases: (1) Calculate the privacy budget for the round $t$ in lines 1–3. (2) Obfuscate the input data $x_k^t$ of user $k$ according to Eq.~\ref{PGEM}. (3) User $k$ utilizes the obfuscated data $\hat{x_k^t}$ for local updates, thereby deriving the model parameters $\hat{w_k^t}$, which are subsequently uploaded to the server.

\section{PERFORMANCE EVALUATION}\label{PERFORMANCE EVALUATION}

\subsection{Experimental Setup}
\textbf{Datasets.} We evaluate the performance of attack and defense strategies on three real-world location datasets: NYCB\footnote[1]{https://www.kaggle.com/datasets/stoney71/new-york-city-transport-statistics}, Tokyo\footnote[2]{https://sites.google.com/site/yangdingqi/home/foursquare-dataset}, and Gowalla\footnote[3]{http://snap.stanford.edu/data/loc-gowalla.html}. The statistics of each dataset are shown in Table~\ref{tab: Basic dataset statistics.}. For each dataset, we randomly select 100 users to participate in federated training and select trajectory points in roughly 10-minute increments.

\begin{table}[]
\caption{ Basic dataset statistics.}
    \centering
    \begin{tabular}{cccc}
    \hline
                 &  NYCB   \qquad  &Tokyo \qquad    & Gowalla\\    \hline
     $\#$users      &   1064   &2293     &53008    \\
     $\#$locations  &   5136   &7872     &121944  \\  
     $\#$check-ins  &   147939  & 447571 & 3302414\\ \hline
    \end{tabular}
    \label{tab: Basic dataset statistics.}
    \label{Table 2}
\end{table}

\subsubsection{Metrics.} We use the attack success rate (ASR) to evaluate the impact of the reconstruction attack, where ASR represents the percentage of accurately reconstructed training data relative to the entire training data~\cite{wei2020framework}. We set the attack to be successful when the Euclidean distance between the reconstructed and true locations is less than 500 meters. We employ attack iteration (AIT) to measure the cost of the attack, where AIT is the number of attack iterations required for a successful attack~\cite{wei2020framework}. To evaluate the efficacy of defense strategies, we adopt the top-$k$ recall rate, recall@5, as a metric to assess the predictive performance.

\subsubsection{Attack Methods.} In addition to ST-GIA, we consider another five attacks that can be applied to spatiotemporal federated learning directly or with simple modifications, including DLG~\cite{zhu2019deep}, iDLG~\cite{zhao2020idlg}, InvGrad~\cite{geiping2020inverting}, CPL~\cite{wei2020framework}, and SAPAG~\cite{wang2020sapag}. See Table~\ref{comparison_state_of_art} for details.

\subsubsection{Defense Strategies.} We compare three defense methods based on differential privacy, including DPSGD~\cite{abadi2016deep}, GeoI~\cite{andres2013geo}, and GeoGI~\cite{takagi2020geo}. They are widely used to protect privacy in spatio-temporal federated learning.

\subsection{Results on Attack Methods}
\textbf{Comparison of Different Attack Methods.}
We first compare the performance of different attacks on three datasets in Table~\ref{table: global comparion1} and Table~\ref{table: global comparion2}. We perform these attacks under different global training rounds and report their attack success rates. We observe that the success rate of all methods tends to decrease as the number of global rounds increases. This is because these attacks require information leaked from the gradient to reconstruct the user's location. However, as the global model convergence, the gradient progressively reveals less information, leading to increased reconstruction errors. Further, we find that ST-GIA consistently outperforms others in the same round. This superior performance can be attributed to our customized attack design for spatiotemporal data, which verified the effectiveness of the proposed attack method. We also observe that InvGrad always shows the worst results, frequently achieving an attack success rate of 0. This poor performance is due to its use of cosine loss as the distance function, resulting in excessively slow convergence. In our experiments, it is noted that InvGrad often required over 20,000 iterations to achieve relatively accurate results, while the maximum attack round is limited to 200. Due to space limitations, we only show the results on two datasets. We observe similar conclusions on the Tokyo dataset.

\begin{table}[thbp]
\caption{\small Comparison of different attack methods (NYCB dataset).}
\centering
 {
\small{
\begin{tabular}{|c|p{1.2cm}|p{1.2cm}|p{1.2cm}|p{1.2cm}|p{1.2cm}|p{1.2cm}|p{1.2cm}|}
\hline

\multicolumn{2}{|c|}{\diagbox{Model}{Round}}  & 1   & 10  & 20  & 30   & 40 & 50 \\ \hline
\multirow{1}{*}{DLG}       & ASR      & 0.781  & 0.687        & 0.531        & 0.464          & 0.332     & 0.269    \\ \cline{2-8}   \hline
\multirow{1}{*}{iDLG}      & ASR      & 0.531     & 0.472          & 0.418      & 0.366           & 0.301 &  0.245   \\ \hline
\multirow{1}{*}{InvGrad}   & ASR      & 0.031   &    0.015         & 0.006      & 0.001          & 0    &     0   \\ \hline
\multirow{1}{*}{CPL}       & ASR      & 0.575   & 0.523            & 0.452         & 0.395            & 0.321&     0.281          \\ \hline
\multirow{1}{*}{SAPAG}     & ASR      & 0.406   &  0.344           &  0.313        &   0.312           &    0.250         & 0.214    \\ \hline   
\multirow{1}{*}{\textbf{ST-GIA}}     & ASR      &  0.895   & 0.825            &  0.761        &   0.709            &   0.593          &  0.425    \\ \hline           
\end{tabular}
}}
\label{table: global comparion1}
\end{table}

\begin{table}[thbp]
\caption{\small Comparison of different attack methods (Gowalla dataset).}
\centering
 {
\small{
\begin{tabular}{|c|p{1.2cm}|p{1.2cm}|p{1.2cm}|p{1.2cm}|p{1.2cm}|p{1.2cm}|p{1.2cm}|}
\hline

\multicolumn{2}{|c|}{\diagbox{Model}{Round}}  & 1   & 10  & 20  & 30   & 40 & 50 \\ \hline
\multirow{1}{*}{DLG}       & ASR      & 0.281  & 0.252       & 0.218        & 0.173        & 0.191     & 0.153    \\ \cline{2-8}   \hline
\multirow{1}{*}{iDLG}      & ASR      & 0.217    & 0.187          & 0.156      & 0.125           & 0.121 &  0.117   \\ \hline
\multirow{1}{*}{InvGrad}   & ASR      & 0   &    0        & 0     & 0         & 0    &     0   \\ \hline
\multirow{1}{*}{CPL}       & ASR      & 0.224   & 0.217           & 0.178       & 0.153            & 0.147  &     0.141          \\ \hline
\multirow{1}{*}{SAPAG}     & ASR      & 0.125   &  0.141           &  0.005        &   0.004           &    0.002         & 0    \\ \hline   
\multirow{1}{*}{\textbf{ST-GIA}}     & ASR      &  0.562   & 0.531            &  0.312        &   0.250            &   0.226          &  0.187    \\ \hline           
\end{tabular}
}
}
\label{table: global comparion2}
\end{table}

\subsection{Ablation Studies} As introduced in Section~\ref{Spatiotemporal Gradient Inversion Attack}, ST-GIA enhances the effectiveness of the attack in three ways: initialization based on spatiotemporal features, mapping, and calibration with multiple recoveries. In this section, we evaluate the effects of these three components on the attack results.

\begin{figure}[htbp] 
  \centering           
  \subfigure[Initialization] {\includegraphics[width=0.31\textwidth]{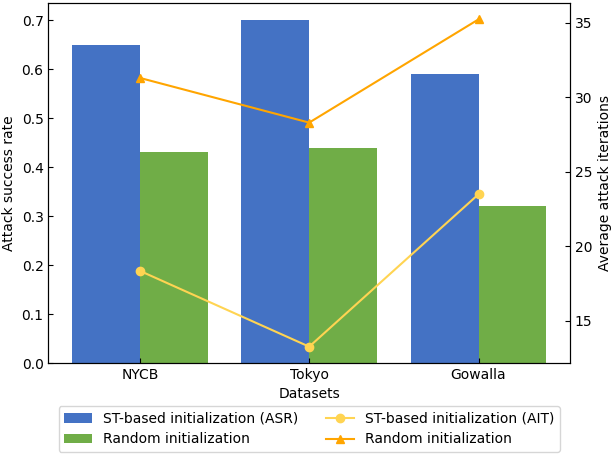}}
  \subfigure[Mapping] {\includegraphics[width=0.31\textwidth]{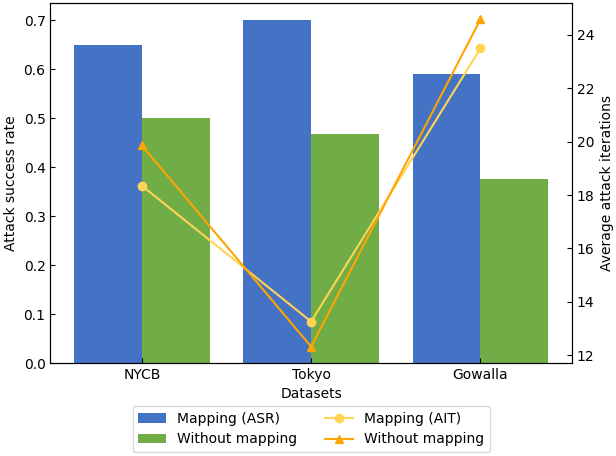}}
   \subfigure[Calibration] {\includegraphics[width=0.31\textwidth]{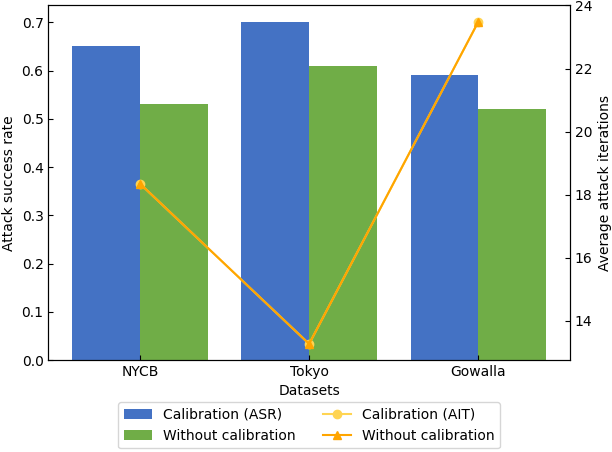}}
  \caption{Ablation studies.}    
  \label{Ablation experiment}    
\end{figure}

\textbf{Impacts of Initialization.}
As shown in Fig.~\ref{Ablation experiment}(a), to evaluate the impact of the initialization component of the proposed attack method on the attack success rate of the three datasets, we employ random and spatiotemporal feature-based initialization methods to initialize the dummy inputs in each round respectively, and all other conditions remain the same. Compared with random initialization, initialization based on spatiotemporal features can significantly increase the attack success rate (at least 20\%). The reason behind this improvement is that poor initialization impedes the attack model's convergence and potentially escalates the associated overhead. Conversely, our initialization method can take advantage of the continuity of user mobility to make educated guesses.

\textbf{Impacts of Mapping.} To evaluate the impact of the mapping component in our proposed attack algorithm, we perform experiments on three datasets. The results shown in Fig.~\ref{Ablation experiment}(b) show that implementing straightforward mapping operations can substantially enhance the attack success rate. This increase in performance is primarily due to the mapping's role in facilitating the convergence of the attack model and ensuring that the results of each iteration fall within a plausible range.  Nevertheless, we also observe that this operation has some negative optimizations over the attack iteration. This issue arises because the mapping operation may inadvertently direct the intermediate results away from the optimal path. However, this drawback is considered acceptable, as it exerts minimal influence on the overall cost of the attack.

\textbf{Impacts of Calibration.} Fig.~\ref{Ablation experiment}(c) shows the impact of calibration components on our proposed attack algorithm. We can observe that attack accuracy can be improved to some extent by calibrating the final results from multiple recoveries. In addition, the calibration process has no impact on the attack iterations, as it is only a post-processing of the attack results.

\subsection{Results on Defense Strategies}
In this set of experiments, we study the effectiveness of various defense strategies against gradient inversion attacks by analyzing the relationship between model prediction accuracy and attack success rate. 
\begin{table}[]
\caption{ The attack success rates of various defense strategies under different privacy budgets $\epsilon$. (Left: NYCB dataset, Right: Gowalla dataset)}
\begin{minipage}[c]{0.30\linewidth}
    \centering
    \setlength{\tabcolsep}{1.5mm}{\scalebox{0.9}{
    \begin{tabular}{cccccc}
    \hline 
    $\epsilon$       &  1    &5  &10  & 20 & 50 \\  \hline
     DPSGD    &   0.06   & 0.14    & 0.23 &0.40 &0.51  \\
     GeoI   &   0.08   & 0.17  & 0.26 & 0.38 &0.54\\  
     GeoGI  & 0.11   &0.20  &0.28 &0.44&0.56\\ 
     Ours    & 0.09 &0.15   & 0.23   & 0.34 & 0.48 \\ \hline
    \end{tabular}
    }}
\end{minipage}
\hspace{2.6cm}    
\begin{minipage}[c]{0.30\linewidth}
    	\setlength{\tabcolsep}{1.5mm} {\scalebox{0.9}{
    \begin{tabular}{cccccc}
    \hline 
      $\epsilon$       &  1    &5  &10  & 20 & 50 \\  \hline
     DPSGD    &  0.04    &0.11  &0.19  & 0.21 & 0.37 \\
     GeoI   &   0.09   & 0.12  & 0.22 & 0.34 &0.41\\  
     GeoGI  & 0.11   &0.14  &0.26 &0.30 &0.39\\ 
     Ours    & 0.02 &0.10   & 0.17   & 0.24 & 0.35 \\ \hline
    \end{tabular}
    }
    }
    \label{different privacy budgets}
\end{minipage}
\end{table}

\textbf{The Effects of Defense Strategies on ASR.} Table~\ref{different privacy budgets} shows the attack success rates of various defense strategies under different privacy budgets $\epsilon$ on the NYKB and Gowalla datasets. We observe that in all cases, the gradient inversion attack is largely mitigated at some cost of accuracy if we add sufficient noise. This result suggests that all defense strategies evaluated can mitigate gradient inversion attacks to some extent. We also observe that in most cases, when the privacy budget is the same, the proposed defense strategy has a lower ASR, proving it is better resistant to gradient inversion attacks. These outcomes validate the effectiveness of the proposed defense strategy.

\begin{figure}[htbp] 
  \centering           
  \subfigure[Results on NYCB dataset] {\includegraphics[width=5cm]{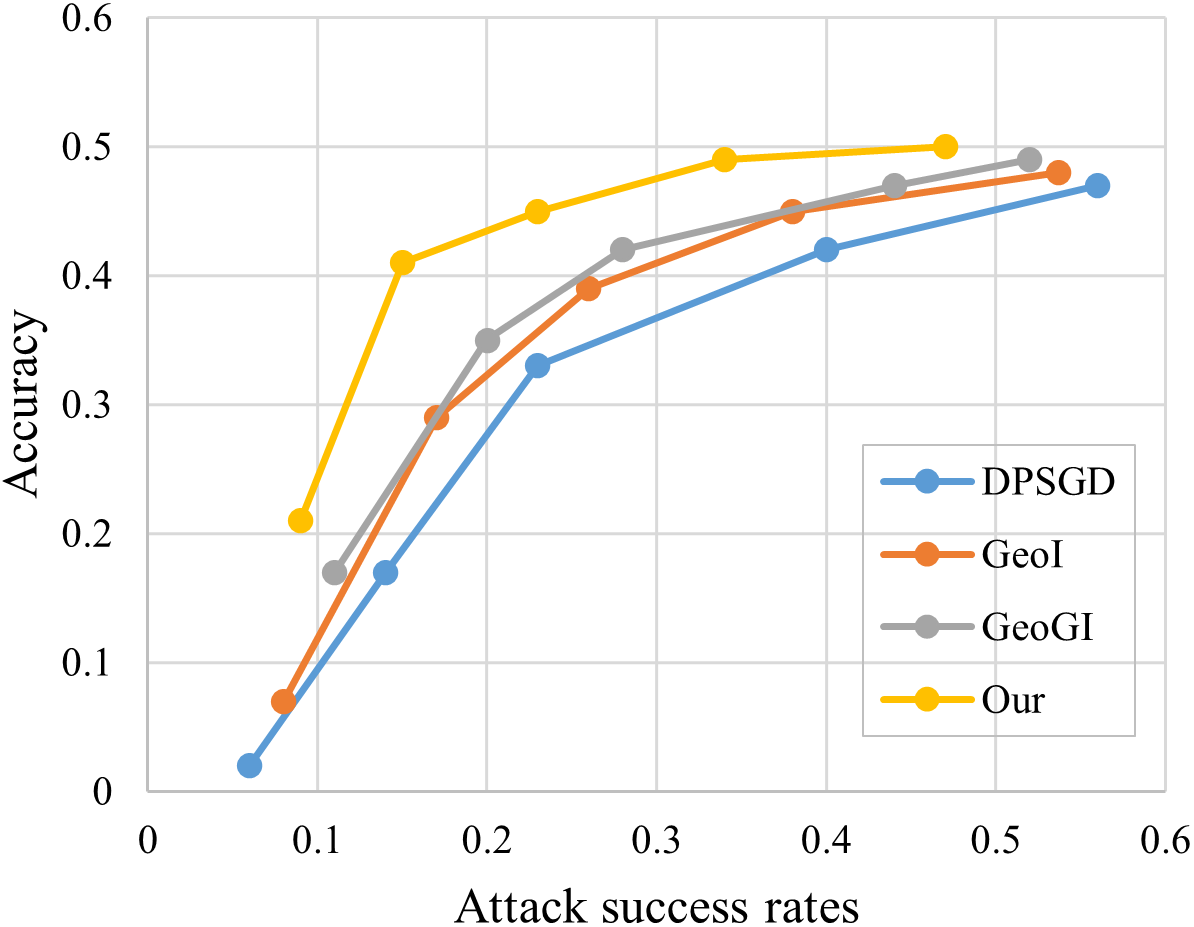}}
  \hspace{1cm}  
  \subfigure[Results on Gowalla dataset] {\includegraphics[width=5cm]{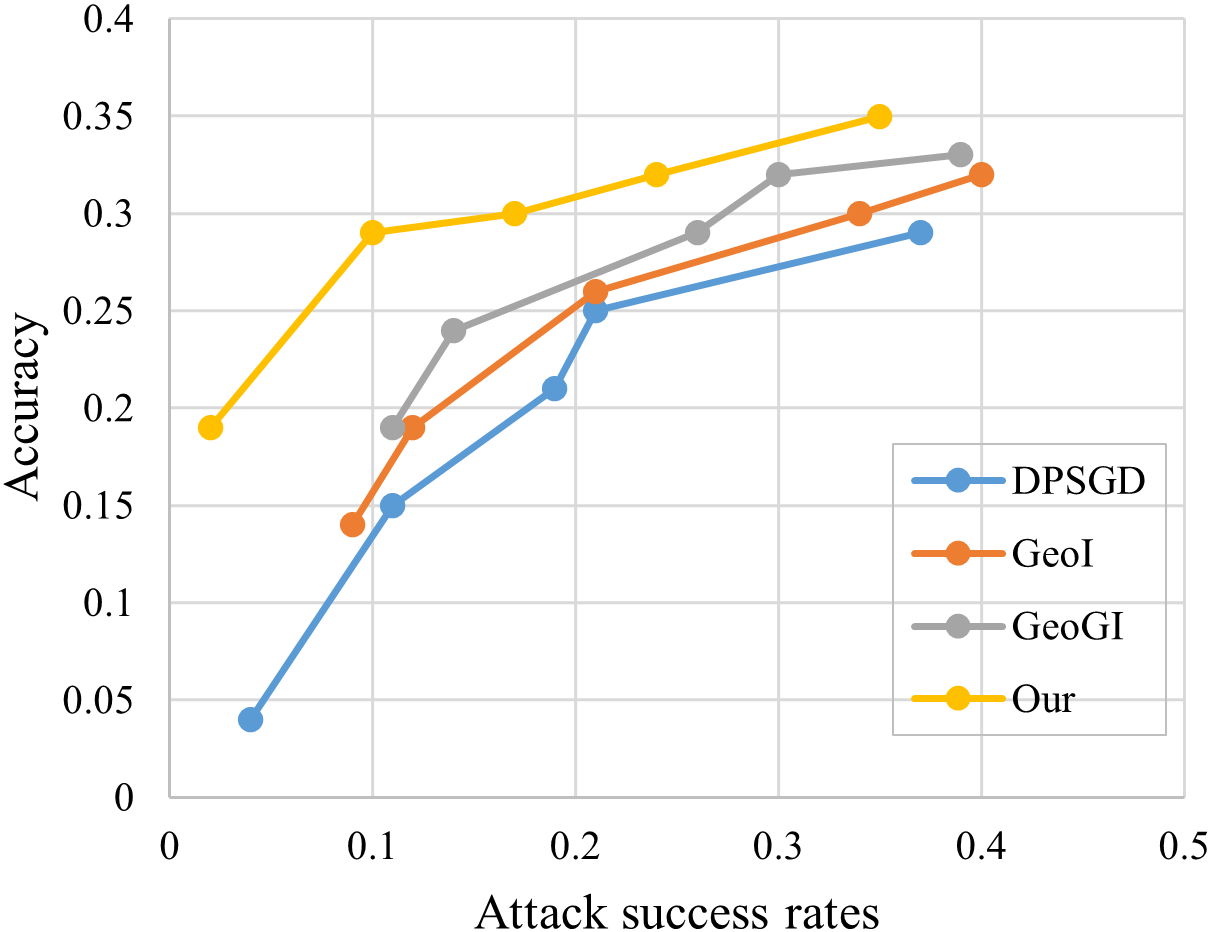}}
  \caption{The relationship between prediction accuracy and attack success rate.}    
  \label{Privacy-utility trade-off}    
\end{figure}
\textbf{The Trade-off Between Attack Risk and Prediction Accuracy.} Fig.~\ref{Privacy-utility trade-off} shows the relationship between prediction accuracy and attack success rate. We also first analyze the NYCB dataset. As shown in Fig.~\ref{Privacy-utility trade-off}(a), when facing equivalent privacy risks  (that is, the attack success rate is the same), the model prediction accuracy of the proposed defense strategy always remains the highest. This shows that our defense strategy achieves the best trade-off between privacy and utility. We can also observe that DPSGD always exhibits the worst performance in most cases. This phenomenon is because other methods are specifically designed for location privacy, but DPSGD represents a more general privacy-preserving method in deep learning. Fig.~\ref{Privacy-utility trade-off}(b) shows similar conclusions on the Gollow dataset.  In summary, our method can achieve better performance with a lower risk of gradient inversion attacks.

\section{Related Work}
\subsubsection{Spatiotemporal Federated Learning.}
Human mobility prediction is one of the most popular tasks in spatiotemporal federated learning and has been extensively studied in recent years~\cite{fan2019decentralized,feng2020pmf,li2020predicting,wang2022location,zhang2023federated}. For example, Li et al.~\cite{li2020predicting} developed a spatial-temporal self-attention network to integrate spatiotemporal information for enhanced location prediction. Feng et al.~\cite{feng2020pmf} introduced a privacy-preserving mobility prediction framework PMF using federated learning, which exhibited notable performance compared to centralized models. Fan et al.~\cite{fan2019decentralized} designed a decentralized attention-based personalized human mobility prediction model and implemented pre-training strategies to expedite the federated learning process. However, malicious attackers may steal private information from shared gradients through gradient inversion attacks.

\subsubsection{Gradient Inversion Attacks.} Gradient inversion attacks were first proposed in 2019 and have been extensively studied in the field of image and text recovery~\cite{deng2021tag,geiping2020inverting,geng2023improved,wang2020sapag,wei2020framework,zhao2020idlg,zhu2019deep}. For example, DLG~\cite{zhu2019deep} uses Euclidean distance as the loss function to measure the distance between generated gradients and ground-truth gradients. 
InvGrad~\cite{geiping2020inverting} chooses to minimize cosine, as opposed to Euclidean loss, to match the direction, not the magnitude, of the true gradient. CPL~\cite{wei2020framework} uses $l_2$ distance and label-based regularization to increase the attack capability further. SAPAG~\cite{wang2020sapag} uses a weighted Gaussian kernel as the distance metric. However, current works focus on reconstructing training images or
texts and have yet to be validated in spatiotemporal federated learning.

\section{Conclusion}
This paper studies gradient inversion attacks in spatiotemporal federated learning for the first time. We introduce a novel attack method, ST-GIA, specifically designed for spatiotemporal federated learning, which can effectively recover accurate original locations from gradients. Subsequently, we develop an adaptive differential privacy method to mitigate gradient inversion attacks. Extensive experimental results confirm the effectiveness of the proposed attack methods and defense strategies. In the future, we will explore more spatiotemporal federated learning scenarios, such as traffic flow prediction.

\begin{credits}
\subsubsection{\ackname} This work is supported in part by JST CREST (No. JPMJCR21M2); in part by JSPS KAKENHI (No. JP22H00521, JP22H03595, JP21K19767); in part by JST/NSF Joint Research SICORP (No. JPMJSC2107); in part by the National Natural Science Foundation of China (No. 62220106004, 61972308).
\end{credits}

\bibliographystyle{splncs04}
\bibliography{refs}

\end{document}